\documentstyle[12pt]{article}
\def\lo{\langle 0 |}
\def\ro{ | 0 \rangle }

\def\gmmu{\gamma _{\mu}}

\def\atop{ \frac{ \alpha_{s}}{4 \pi} G_{\mu \nu}
 \tilde{G}_{\mu \nu} }

\def\fc{ f_{\eta'}^{(c)} }
\def\gmf{\gamma _{5}}

\def\la{\langle }
\def\ra{ \rangle }
\def\el{ \langle \eta'| }
\def\er{ | \eta' \rangle}

\newcommand{\beq}{\begin{equation}}
\newcommand{\eeq}{\end{equation}}
\newcommand{\bea}{\begin{eqnarray}}
\newcommand{\eea}{\end{eqnarray}}

\begin{document}
                                        \begin{titlepage}
\begin{flushright}
hep-ph/9705251
\end{flushright}
\vskip1.8cm
\begin{center}
{\LARGE
 Why  is the $ B \rightarrow \eta'X $ decay 
width so large? 
   } 
\vskip1.0cm
            
\vskip1.5cm
 {\Large Igor~Halperin} 
and 
{\Large Ariel~Zhitnitsky}
\vskip0.2cm
        Physics and Astronomy Department \\
        University of British Columbia \\
        6224 Agriculture Road, Vancouver, BC V6T 1Z1, Canada \\  
     {\small e-mail: 
higor@physics.ubc.ca \\
arz@physics.ubc.ca }\\
\vskip1.5cm
{\Large Abstract:\\}
\end{center}
\parbox[t]{\textwidth}{
New mechanism for the observed inclusive $ B \rightarrow \eta'X $ decay
is suggested. We argue that the dominant contribution
to this amplitude is due to the Cabbibo favored $b\rightarrow\bar{c}cs$ 
process followed by the transition $\bar{c}c\rightarrow\eta'$.
A large  magnitude of  the ``intrinsic charm" component 
of $\eta'$ is of critical importance in our approach. 
 Our results are consistent with an unexpectedly 
large $ Br( B \rightarrow \eta'+X) \sim  10^{-3} $  
recently announced by CLEO.
We stress the uniqueness of this channel for $0^{-+}$ gluonia search.}

\vspace{1.0cm}

                                                \end{titlepage}

{\bf 1.}Recently CLEO has reported\cite{CLEO} a very 
large branching ratio for the 
inclusive production of $\eta'$ :
\beq
\label{Kim}
Br(B \rightarrow \eta'+X; 2.2 GeV\leq E_{\eta'}\leq 2.7 GeV) = (7.5
 \pm 1.5 \pm 1.1) \cdot 10^{-4},
\eeq
where quoted above result 
contains an acceptance cut intended to reduce the background 
from events with charmed mesons.
To get a feeling of how large this number is, we 
present for comparison a   branching ratio for the 
inclusive production of $J/\psi$ meson \cite{PDF}:
\beq
\label{psi}
Br(B \rightarrow J/\psi ( direct)+X ) = (8.0
 \pm 0.8) \cdot 10^{-3}
\eeq
This process is due to the Cabbibo favored 
$b\rightarrow\bar{c}cs$ decay which is largest possible 
amplitude without   charmed hadrons (like $D, D_s, \Lambda_c...$)
in the final state.
The comparison of these two numbers shows that
the amplitude of process (\ref{Kim})  
is only by a factor of $3$ less than the most  Cabbibo favored 
amplitude $b\rightarrow\bar{c}cs\rightarrow  J/\psi s$.
It is clear that the data (\ref{Kim}) is in severe contradiction 
with a standard view of the process at the 
quark level as a decay of the $b$-quark into 
 light quarks which could be naively suggested  
 keeping in mind the standard picture of   
 $ \eta' $ as  a SU(3) singlet
meson made of the $u-$, $d-$ and $s-$quarks. In this picture 
the decay (\ref{Kim}) must be proportional to the Cabbibo suppression
factor $V_{ub}$, and therefore the standard 
approach has no chance to explain 
the data (\ref{Kim}), see below for more detail.
Once this fact is realized,
we should look around and ask
the question: where does the $\eta'$ come from? We remind
that it has been known
\cite{Witten,Ven}
for a long time that $\eta'$ is a very special meson.
It is so special that physicists repeatedly organize 
 workshops with the word $\eta'$ on the title \cite{eta}.
The question addressed there can be formulated in the following way:
What kind of experiments should be  designed
to demonstrate the uniqueness of $\eta'$?

The aim of this letter is to argue
that one of the crucial
experiments establishing the uniqueness
of $\eta'$ was not only designed, but rather
it was already successfully completed (\ref{Kim})! 
The reason why the $b\rightarrow\eta'$ transition is so
unique for the study of $\eta'$
can be explained in simple terms as follows.
We claim that the $\eta'$ production is
due to the Cabbibo favored $b\rightarrow\bar{c}cs$ 
process followed by the transition 
$\bar{c}c\rightarrow\eta'$.
Each step here is under theoretical control:
the Cabbibo favored $b\rightarrow\bar{c}cs$ 
transition is a prerogative of the weak interactions
 where pair $\bar{c}c$ is created at    small distances 
($x\sim M_W^{-1}$) and 
the amplitude  is proportional to $V_{cb}$. We have nothing new 
to say
at this stage. The second stage is more interesting
and related to the transition $\bar{c}c \rightarrow gluons 
\rightarrow \eta' $.
Due to the fact that the $c$-quark could be considered 
as a heavy particle, one can  
perform the $1/m_c$ expansion reducing the original problem to 
the problem of the gluon content of $\eta'$. Therefore, 
in the $B\rightarrow\eta'$ decay  we have a new 
local gluon source
which has never been available before\footnote{ To be 
more precise,
we should mention that a gluon source is also provided by the 
radiative $J/ \psi\rightarrow\gamma gg$ decays. However, 
gluons are created there not locally, but rather have a very broad
momentum distribution. It makes a theoretical analysis much 
more complicated
and involved, and therefore corresponding predictions become
more ambiguous. What is more important, the mass of a
pseudoscalar gluonium is expected to be very large 
($m_{0^-}> 2.5 \; GeV $ \cite{NSVZ})
and a corresponding gluonium production might be not available 
in $J/ \psi$ 
radiative decays.}.
We should stress that our mechanism is very different from one 
proposed recently \cite{Soni}, and distinctions between them
will be explained below.

{\bf2.} We would like to start our presentation 
with an estimate of the 
$ B \rightarrow \eta' +X$ decay width assuming that the
 $ \eta' $ 
is made exclusively of light quarks. To this end,
it is convenient to consider the following ratio
for two pseudoscalar particles, $\eta'$ and $\eta_c(1S)$ :
\bea
\label{1}
\frac{\Gamma(B \rightarrow  \eta'+X)}{\Gamma(B \rightarrow 
 \eta_c(1S)+X)} \sim(\frac{V_{bu}}{V_{bc}})^2\frac{ | \el 
\bar{u} \gmmu \gmf u \ro \la 
X| \bar{d} \gmmu(1+\gmf) b | B \ra |^2}{  | \la \eta_c| 
\bar{c} \gmmu\gmf c \ro \la 
X | \bar{s} \gmmu(1+\gmf) b | B \ra |^2} 
\frac{\Omega_{B \rightarrow  \eta'+X}}{
 \Omega_{B \rightarrow \eta_c+X}} \\   \nonumber
\sim
 \frac{1}{3} (\frac{V_{bu}}{V_{bc}})^2\left(\frac{
f_{\eta'}}{ f_{\eta_c}} \right)^2   
\left(\frac{1-m_{\eta'}^2/m_b^2}{1-m_{\eta_c}^2/m_b^2}\right)^2
\sim  3\cdot 10^{-4}
\eea
Here $\Omega_{B \rightarrow  \eta'+X}$ and 
$\Omega_{B \rightarrow \eta_c+X}$
are the corresponding phase volumes for two inclusive decays;
$(\frac{V_{bu}}{V_{bc}})\simeq 0.08$. The matrix element 
$ \la \eta'(p) | 
\bar{u} \gmmu \gmf u \ro =\frac{-i}{\sqrt{3}}f_{\eta'}p_{\mu}
\simeq  (0.5\div 0.8)\frac{-i}{\sqrt{3}}f_{\pi}p_{\mu}$ is 
known numerically
from \cite{NSVZ}. We define the corresponding $\eta_c$
 matrix element in a similar way to the $ \eta'$:
\beq
\label{eta_c}
 \la \eta_c(p) | 
\bar{c} \gmmu\gmf c\ro =-if_{\eta_c}p_{\mu}
\eeq
 This matrix element
can be estimated from the   $\eta_c
\rightarrow\gamma\gamma$ decay:
\beq
\label{2}
\Gamma (\eta_c(1S)
\rightarrow\gamma\gamma)=\frac{4(4\pi\alpha)^2 
f_{\eta_c}^2}{81\pi m_{\eta_c}}
=(7.5_{-1.4}^{+
1.6}) \; KeV ~~\cite{PDF},~~~~ \Rightarrow 
f_{\eta_c}\simeq 400 \; MeV
\eeq
We used the standard nonrelativistic approach in the  derivation of 
Eq.(\ref{2}).
  We should note that in the nonrelativistic model
$f_{\eta_c}$ determines the value of the wave function 
(WF) at the origin. It is related
 to the standard  nonrelativistic WF $R_s(0)$ \cite{Rep} as follows:
\beq
\label{3}
f_{\eta_c}^2=\frac{3}{2\pi m_c}|R_s(0)|^2
 \eeq

To make a prediction for 
$\Gamma(B \rightarrow  \eta'+X)$ from Eq.(\ref{1}),
we need to know $\Gamma(B \rightarrow 
 \eta_c(1S)+X)$ which, unfortunately, is not presently available.
However, as we will see in a moment, $\Gamma(B \rightarrow 
 \eta_c(1S)+X)\simeq 0.6 \cdot\Gamma(B \rightarrow 
 J/\psi+X)$. The latter number is well 
known (\ref{psi}). Therefore, the standard mechanism
yields a very small contribution in comparison 
with the data (\ref{Kim}):
 \beq
\label{4}
Br(B \rightarrow  \eta'+X)_{standard}\sim 1.5\cdot 10^{-6}
\eeq
 We should mention  that the factorization procedure used in 
the estimate (\ref{1}) does
not work well. A  phase factor 
introduced into this formula  is also a
rough simplification: in reality, an inclusive spectrum
is much more complicated function than 
a simple factor $ \Omega_{B \rightarrow  \eta'+X}$
obtained  as a result of two-particle decay of a colorful heavy quark
$b\rightarrow\eta'(\eta_c)+d(s)$ instead of the physical $B$ meson.
  Besides,
 gluon corrections to the Wilson coefficients in front of the  operators 
containing $\bar{c} c$ quarks  (denominator in (\ref{1}))
or light quarks (numerator in (\ref{1}))  
also change the estimate (\ref{1}). 
However, it is obvious that all these effects due to a 
non-factorizability, gluon corrections, 
  as well as  
$ O(1/m_b , 1/N) $ terms omitted in (\ref{1}),
cannot substantially change our estimate. 
We therefore conclude that the image of the $ \eta' $ meson 
as the
SU(3) singlet quark state made exclusively 
of the $ u,d,s $ quarks is not adequate 
to the problem at hand.
It is easy to see that the small  value for the ratio (\ref{1}) is 
a consequence of a small residue of the $ \eta' $
supplemented with the Cabbibo suppression of the 
$ b \rightarrow u $ transition. 

To conclude the discussion of the standard approach
to the $ B \rightarrow X \eta' $ decay, we should
estimate 
$Br(B \rightarrow 
 \eta_c(1S)+X)$ which was not yet measured, but was a relevant element
in our calculations (\ref{1}).
To this end, consider the following ratio 
\bea
\label{5}
\frac{\Gamma(B \rightarrow  \eta_c(1S)+X)}{\Gamma(B \rightarrow 
 J/\psi+X)} \sim \frac{ |\la \eta_c|\bar{c} \gmmu \gmf c\ro \la 
X| \bar{s} \gmmu(1+\gmf) b | B \ra |^2}{  | \la J/\psi| 
\bar{c} \gmmu c \ro \la 
X | \bar{s} \gmmu(1+\gmf) b | B \ra |^2} 
\frac{\Omega_{B \rightarrow  \eta_c+X}}{
 \Omega_{B \rightarrow J/\psi+X}} \\   \nonumber
\sim
\frac{1}{1+2 m_{J/\psi}^2/m_b^2} \cdot   \left(\frac{
f_{\eta_c}}{ f_{\psi}} \right)^2 
\sim   0.6
\eea
Here we introduced the constant $f_{\psi}$ 
defined by the following matrix element:
\beq
\label{6}
\la J/\psi| 
\bar{c} \gmmu c \ro   =\epsilon_{\mu}f_{\psi}m_{\psi}
\eeq
The definition of   $f_{\psi}$ is similar to   the definition of 
 $f_{\eta_c}$ introduced before (\ref{eta_c}).  In the 
nonrelativistic limit
these residues are equal $f_{\psi}=f_{\eta_c}$, and both 
can be expressed  
in terms of the number $R_s(0)$, see (\ref{3}). Such an equality
is a consequence of the fact that $J/\psi$ and $\eta_c$
mesons are the two states ($^3S_1$ and $^1S_0$ ) of the 
same $\bar{c}c$-system 
with the same quantum numbers   $|n=1,l=0\ra$. One can estimate
$f_{\psi}$ independently from $J/\psi\rightarrow e^+e^-$ decay:
\beq
\label{7}
\Gamma (J/\psi 
\rightarrow e^+e^-)=\frac{(4\pi\alpha)^2 
f_{\psi}^2}{27\pi m_{J/\psi}}
=(5.26\pm 0.37) \; KeV ~~\cite{PDF},~~~~ \Rightarrow 
f_{\psi}\simeq 400 \; MeV
\eeq
with the result that experimentally the ratio 
$f_{\psi}\simeq f_{\eta_c}$ 
is fulfilled within the errors.
(As our purpose here is the order of magnitude estimate
(\ref{5}), we neglect all relativistic corrections \cite{rel}
which could change the relation $ f_{\eta_c} = f_{\psi} $.)
We also note that in the limit $m_b\rightarrow
\infty$ only the longitudinal polarization of $J/\psi$ meson
contributes the
decay, see e.g. \cite{Brod1}. In this case
$\epsilon_{\mu} m_{\psi}\rightarrow p_{\mu}$,  
and therefore the matrix elements for 
longitudinally polarised $J/\psi$ meson
(\ref{6}) and $\eta_c$ meson (\ref{eta_c}) are equal,
and the ratio (\ref{5})
should be close to one: $(\frac{
f_{\eta_c}}{ f_{\psi}})^2\simeq 1$. 
In reality, $m_b$ is not much heavier than
$J/\psi$, and thus the contribution of  
two transverse polarizations of $J/\psi$
  is not suppressed numerically, and the correction factor 
due to the transverse polarizations is explicitly taken into 
account in (\ref{5}).

Our last remark regarding Eq.(\ref{5}) is that  
it is very important that this ratio is not 
sensitive to the problems 
of non-factorizability, gluon corrections and 
many others we mentioned (and did not mention)  earlier.
This is because all uncertainties related to those
problems are cancelled out in the ratio (\ref{5}).

{\bf 3.} In view of the failure of the 
standard approach to the $ B \rightarrow \eta' +X$ decay 
which treats the  $ \eta' $  
as the
SU(3) singlet quark state made exclusively 
of the $ u,d,s $ quarks,   we suggest an 
alternative mechanism for the $ B \rightarrow\eta'+X $ decay 
which is specific to the uniqueness of the  
  $ \eta' $. It has been known \cite{Witten,Ven},
that the $ \eta' $ is {\bf  a messenger between two worlds}:
the world of light hadrons
and a less studied world  of gluonia.  In other words, it 
is a very special meson strongly coupled to gluons.  
We suggest the following picture for the process of interest:  
 the $ b \rightarrow c\bar{c} s $ 
decay is followed by the conversion of the $c\bar{c}$-pair into
 the $ \eta'$
\footnote{The relevance of the process $ b \rightarrow c \bar{c} s 
\rightarrow light \; hadrons $ was discussed earlier 
\cite{Dun} in connection
with the problem of semileptonic branching ratio.}.  
This means that the matrix element 
\beq
\label{9}
\lo \bar{c} \gmmu \gmf c | \eta'(p) \ra = i \fc p_{\mu}
\eeq
is non-zero due to the $ c \bar{c}\rightarrow  gluons $ transition. 
Of course,
since one deals here with virtual c-quarks, this matrix 
element is 
suppressed by the $ 1/m_{c}^2 $ However,
 the c-quark is not very heavy, and the suppression
$ 1/m_{c}^2 $ is not large numerically. At the same time,
 the Cabbibo
enhancement of the $ b \rightarrow c $ transition in
 comparison
to $ b \rightarrow u $ is a much more important factor 
which makes this mechanism work.

In our recent  paper \cite{1} we
estimated the matrix element (\ref{9})  
 using a combination of the 
Operator Product Expansion technique, large $ N $ approach and
QCD low energy theorems.
The final formula reads 
 \beq 
\label{13}
\fc = - \frac{
\lo g^3 f^{abc} G_{\mu \nu}^a \tilde{G}_{\nu \alpha}^b 
G_{\alpha \mu}^c \er }{16 \pi^2 m_{\eta'}^2 m_{c}^2}
+ O \left( \frac{1}{m_{c}^4 } \right) \simeq
 \frac{3}{4 \pi^2 b} \frac{1}{m_c^2} \frac{
\la g^3 G^3 \ra _{YM}}{ \lo \atop \er }  + O \left( \frac{1}{m_{c}^4 }
\right) 
\eeq
Therefore,   
  we have related the residue of the charmed 
axial current into the $ \eta'$ with apparently 
completely unrelated quantity which is the value of 
cubic gluon condensate in pure Yang-Mills theory
(we notice that the matrix element of topological density 
  which appears in
(\ref{13}) is known: $ \lo (\alpha_s/ 4\pi) G \tilde{G}
\er \simeq 0.04 \; GeV^3 $ \cite{NSVZ}).
Using all currently available information 
regarding the vacuum condensate
$\la g^3 G^3 \ra _{YM}$ in gluodynamics,
we have arrived at the numerical estimate 
\beq
\label{10}
\fc  = ( 50 \div 180) \; MeV
\eeq 
Here the  
 uncertainty is mostly due to a poor knowledge of the 
cubic condensate in gluodynamics.
The residue $ \fc $ has also been calculated
 numerically \cite{SZ} within the instanton liquid model, where 
it was found $ \fc = (100 -120) \; MeV $ in agreement with 
(\ref{10}). 
 
In spite of the poor accuracy of our result (\ref{10}), we  
concluded in \cite{1}
that the gluon mechanism seems to be sufficient 
to describe the data for exclusive decay $B\rightarrow\eta'K$.
We came to this conclusion by comparing the 
theoretical prediction (\ref{10}) with an ``experimental"
value of $ \fc $ obtained under assumption that the 
above mechanism exhausts the $ B \rightarrow K \eta' $ decay.
From the numerical estimate $
Br( B \rightarrow K \eta') \simeq 3.92 \cdot 10^{-3} \cdot
( f_{\eta'}^{(c)}/1 \; GeV )^2 $
and the CLEO data \cite{CLEO} 
\beq
\label{Kim1}
Br(B \rightarrow K \eta') = (7.8_{-2.2}^{+
2.7} \pm 1.0) \cdot 10^{-5} \; ,
\eeq
we have found
the ``experimental" value (we use the central value 
of the branching ratio (\ref{Kim1}) )  
\beq
\label{16}
\fc \simeq 140 \; MeV \; \; (``exp")
\eeq 
which is within our estimate of $\fc$ (\ref{10}). Bearing in 
mind that the standard approach to $ B \rightarrow K \eta' $ yields
$Br( B \rightarrow K \eta') \simeq 10^{-7} $ (which is extremely 
small in comparison to (\ref{Kim1})), we concluded that the suggested
mechanism indeed explains the exclusive decay $ 
 B \rightarrow K \eta' $, with a reservation for uncertainty of 
our prediction (\ref{10}).   

Before proceeding with the use of our estimate (\ref{10})
for the inclusive decay $B\rightarrow\eta'+X$,
we would like to make a few comments. The obtained result(\ref{10})
looks very large as 
 it is only a few times smaller than the analogously 
normalized residue 
for $\eta_c$ meson, see (\ref{eta_c},\ref{2}).
At the same time, $\fc$ is a double suppressed amplitude:
it is Zweig rule-violating
and besides contains the $1/m_c^2$ suppression factor.
 Therefore,  presumably, it should be very small. 
In reality it is not.
There are two reasons
for this. First, $m_c^2$ is not very large on hadronic $1\; GeV$ scale. 
Second, and more important,
the Zweig rule itself is badly broken in 
vacuum $0^{\pm}$ channels.  Of course, it is in 
contradiction
with a naive large $N_c$   
counting where a non-diagonal transitions
should be suppressed in comparison with 
a diagonal ones.
However, a more careful analysis \cite{NSVZ,1} reveals
that the large $ N_c $ picture and the breakdown of the 
Zweig rule in fact peacefully co-exist: while the 
large $ N_c $ description is quite accurate for the 
$ \eta' $, an extent to which the Zweig rule is violated
in $ \eta' $ just sufficies to obtain the large residue 
(\ref{10}) (see \cite{1} for more detail). We stress that 
the phenomenon of a breakdown of the Zweig rule 
in vacuum $ 0^{\pm} $ channels is well known and 
understood \cite{NSVZ}, and many phenomenological
examples of corresponding physics have been discussed in the 
literature, see e.g. \cite{NSVZ,arz}. 
The large residue 
$\fc$ (\ref{10}) (which is fundamentally important 
for our estimates)
is another manifestation of the same physics.
One should expect that the intrinsic charm component of 
the $ \eta' $ can show up in a number of physical 
processes. However, in some cases it does not lead to any 
effects. For instance, the $ \eta' \rightarrow 2 \gamma $ 
decay is not influenced by the $c$-quark in  the $ \eta' $
because the heavy quark contribution to the triangle diagram
vanishes when the photons are on-shell \cite{Muel}.

In a number of papers (see e.g. \cite{Soni,AG}), a much smaller
value of the residue $ \fc $ was suggested, $ \fc \simeq 5.8 \; 
MeV $ \cite{AG}. These estimates are based on the constituent
quark model picture  of the $ \eta_c - \eta' $ mixing
\bea 
\label{mix}
| \eta' \ra &=& \frac{1}{\sqrt{3}} | u \bar {u} + d \bar{d}
+ s \bar{s} \ra \cos \theta + | c \bar{c} \ra \sin \theta \; , 
\nonumber \\
| \eta_c \ra &=& - \frac{1}{\sqrt{3}}  | u \bar {u} + d \bar{d}
+ s \bar{s} \ra \sin \theta + | c \bar{c} \ra \cos \theta \; .
\eea
In our opinion, this approach cannot be correct in view of two reasons. 
First, the quantum mechanical mixing (\ref{mix})
implies that all constituent quarks in (\ref{mix}) are 
nearly on-shell (with an accuracy $ \sim \Lambda_{QCD} $).
Clearly, this cannot be the case for the heavy $c$-quark in the $ \eta'$.
Moreover, this picture completely neglects all gluon Fock 
components in the $ | \eta' \ra $ state. In particular, one could
expect that the matrix element $ \lo \alpha_s G \tilde{G} \er $ should 
be very small, while we know from QCD that it is actually large.
The same is true also for higher gluon Fock states in the $ \eta' $,
which are directly related to the residue of interest (\ref{13}).
  
Once the fundamental parameter $\fc$ is fixed, we can  estimate
the  inclusive decay $Br(B \rightarrow   \eta'+X)$.
As before, it is convenient to consider the following ratio
for two pseudoscalar particles, $\eta'$ and $\eta_c(1S)$ :
\bea
\label{17}
\frac{\Gamma(B \rightarrow  \eta'+X)}{\Gamma(B \rightarrow 
 \eta_c+X)} \sim \frac{ | \el \bar{c} \gmmu \gmf c \ro \la 
X| \bar{s} \gmmu(1+\gmf) b | B \ra |^2}{  | \la \eta_c| 
\bar{c} \gmmu\gmf c \ro \la 
X | \bar{s} \gmmu(1+\gmf) b | B \ra |^2}
\frac{\Omega_{B \rightarrow  \eta'+X}}{
 \Omega_{B \rightarrow \eta_c+X}}  
\sim
  \left(\frac{\fc}
 { f_{\eta_c}} \right)^2   
  \sim 0.12
\eea
Here $\Omega_{B \rightarrow  \eta'+X}$ and 
$\Omega_{B \rightarrow \eta_c+X}$
are the corresponding phase volumes for two inclusive decays. 
As we mentioned earlier, we are quite confident about 
the ratio for $Br(B \rightarrow   \eta_c+X) \simeq 0.6 \cdot
(B \rightarrow 
 J/\psi+X)\sim 5\cdot 10^{-3}$, see (\ref{5}).
Therefore, from (\ref{17}) we   expect 
\beq
\label{18}
Br(B \rightarrow  \eta'+X)\simeq 0.12\cdot
Br(B \rightarrow   \eta_c+X)\simeq 6\cdot 10^{-4} \; , 
\eeq
which is our main result. The obtained number is in a 
good agreement
with the data (\ref{Kim}).
In the course of our estimates (\ref{17}) and (\ref{18})
we have used the ``experimental '' value for $\fc$ (\ref{16}) rather
than our theoretical calculation (\ref{10}) which has a poor
accuracy. 
By doing so, we essentially assumed that our 
mechanism is sufficient 
to describe  
the experimental data for exclusive decay \cite{1}.
The  agreement of (\ref{18}) with the data (\ref{Kim})
for inclusive decay serves as 
 a nontrivial test of the suggested mechanism to 
work also for inclusive decay.

Few words about uncertainties in Eq.(\ref{17}). The most
 important ambiguity in our previous Eq.(\ref{1})
was related to the uncertainty in the Wilson coefficients 
for operators containing 
different flavors. In other words, the standard combination
$ c_1+c_2/N$ which appears in the formulae is very  different for
operators with (or without) charmed quarks. We do not have
such an  uncertainty at all in Eq.(\ref{17}) because we are 
discussing
one and the same operator with the 
charmed quarks for both outgoing particles,
$\eta_c$ and $\eta'$. By the same reasons, we do not have
any uncertainties related to a poor knowledge of $V_{ub}/V_{cb}$
: this factor simply does not appear in (\ref{17}).

The only important uncertainty in (\ref{17}) is related to
our lack of knowledge of $1/m_b$ corrections. 
In the limit $m_b\rightarrow\infty$ these corrections should
disappear. However, in reality they could be large. 
In fact, the phase volume term alone
$\Omega_{B \rightarrow  \eta'+X}/\Omega_{B \rightarrow \eta_c+X}$
calculated on the tree level $b\rightarrow \eta_c(\eta')+s$
would bring in the factor 
$ ( 1-\frac{m_{\eta'}^2}{m_b^2})^2/
(1-\frac{m_{\eta_c}^2}{m_b^2} )^2$ which is $ \sim 1/m_b^2$, but 
is not small numerically. Along with these corrections,
there are corrections related to the
difference in  the spectrum for inclusive
amplitudes $\la  X| \bar{s} \gmmu(1+\gmf) b | B \ra $  
for different $q^2=m_{\eta_c}^2$ or $q^2=m_{\eta'}^2$, respectively.
This effect works in the opposite direction than
the phase volume effects.  Indeed, for any given particle 
(where $X$
is replaced by any definite state 
$K$, $K^{\ast}$...) one should expect the dipole-type
behavior: 
$\la  K(K^{\ast})...| \bar{s} \gmmu(1+\gmf) b | B \ra 
\sim \frac{1}{1-q^2/(m_{b}^{\ast})^2}$ 
with $m_{b}^{\ast}\simeq m_b$. This effect clearly works
to partly cancel 
the phase volume contribution. The net effect of such a 
cancellation
is of order $ \frac{m_{\eta_c}^2}{m_b^2}
\frac{((m_{b}^{\ast})^2-m_b^2)}{m_b^2}
\ll 1$.
A corresponding theoretical analysis   
of all those corrections is still lacking.
 Therefore, we neglect these $1/m_b^2$
corrections altogether. We believe that
such an estimate is much better
than  an alternative  procedure when 
one  takes into account a subset of the corrections
while neglecting all the others
 with the same order of magnitude.
 This is the reason why the estimate (\ref{17}) is so simple and 
reduced to the ratio of corresponding residues only:
$\left( \fc/
  f_{\eta_c} \right)^2   
  \sim 0.12.$
We expect that an accuracy of our estimate (\ref{17}) is rather high
and $O(1/m_b^2)$ corrections cannot considerably 
change our results. 
 
{\bf 4.} The main result of the present letter is expressed 
by the formulae (\ref{17}), (\ref{18}) which agree well
 with the 
data (\ref{Kim}).
 We therefore suggest a mechanism which is responsible
for both decays: exclusive $B\rightarrow \eta' K$ \cite{1} 
as well as  inclusive
$B\rightarrow \eta' +X$ one (\ref{18}). 
The mechanism is based on the Cabbibo favored $b\rightarrow\bar{c}cs$
process followed by the transition $\bar{c}c$ into the $\eta'$.
We believe that all  
similar modes (as, for example,  
 $B\rightarrow \eta' K^{\ast}$) 
will follow the same pattern. Therefore, we expect 
that the   difference between $K$ and $K^{\ast}$ modes 
 appears only in the transition form factors
$\sim \la K (K^{\ast}) | \bar{s} \gmmu(1+\gmf) b | B \ra$. 
At the same time, the part related to
the $\eta'$ remains unchanged.   In this case,  the corresponding
information can be extracted from the analysis of well-known $B 
\rightarrow  J/\psi+K(K^{\ast})$ decays and we arrive at the estimate
\beq
\label{19}
\frac{\Gamma(B \rightarrow  \eta'+K)}{\Gamma(B \rightarrow 
 \eta'+K^{\ast})} \sim \frac{   
| \la K| \bar{s} \gmmu(1+\gmf) b | B \ra |^2}{   
| \la K^{\ast} | \bar{s} \gmmu(1+\gmf) b | B \ra |^2}\sim 
\frac{\Gamma(B \rightarrow  J/\psi+K)}{\Gamma(B \rightarrow 
 J/\psi+K^{\ast})}\sim 0.5
\eeq
We mention that, as before, the main 
uncertainties, related 
to the Wilson coefficients,
are cancelled out in the ratio (\ref{19}). However, pre-asymptotic
in $ 1/m_b $ effects in these decays
(which can be large, see e.g. \cite{Lip}) are not taken into account in
(\ref{19}). Therefore, our prediction (\ref{19}) should be 
considered as an order of magnitude estimate only. 

In closing,  it is important to note that the suggested mechanism
for $B\rightarrow \eta'$ decay is unique to the special nature
of the $ \eta' $, and thus possesses many specific properties 
which can not be explained by any other mechanism. It gives a hope
that future experiments will support (or reject) 
the suggested pattern. 

In particular, we expect that only $0^{-+}$ flavor-singlet mesons (similar 
to $\eta'$) could contribute on the same level as  
 (\ref{Kim}), (\ref{Kim1}). (Of course $\eta$ will 
also appear due to the 
standard mixing with $\eta'$.)
It is in a big contrast, for example, with the mechanism suggested
in \cite{Soni}, where any state 
with any spin and parity ($0^{++},0^{-+},2^{++},4^{++},...$),  
 strongly interacting
 with two gluons should have, in general,  a similar
 branching ratio.  An experience with 
 $J/\psi\rightarrow \gamma gg \rightarrow \gamma + hadrons$ 
decays tells us that many states (among them: $\eta'$, $f_4(2050)$,
$f_2(1270)$, $\rho\rho$ and others) have two-gluon coupling 
 constants comparable with the  magnitude of $gg\rightarrow \eta'$.
Therefore, one could expect that within this scenario
the same states should appear
in $B$ decays also. As we mentioned earlier, we do not expect 
anything but $0^{-+}$-mesons. 

Moreover, we consider this physics
as a new tool for $0^{-+}$ gluonia search. Such a new   gluon 
color-singlet current produced in 
$B\rightarrow\bar{c}c\rightarrow glueball$ decays
has never been available for a study before.
 
To conclude, we should note that the special quantum numbers
   $0^{-+}$ is not the only point which distinguishes  our mechanism 
from  all others. A spectrum in the inclusive decay is also
very unique: in the 
$m_b\rightarrow\infty$ limit, it is given by  the free quark decay result: 
  $b\rightarrow \eta'+s$ with the specific two-particle decay
spectrum. Physical interactions in $B\rightarrow\eta'X$ will smear
this spectrum with a width about $1 \; GeV$, however, even in such
form it will be very distinguishable from predictions
of other mechanisms.
Therefore, we suggest that this uniqueness of the spectrum 
will be helpful for the
$0^{-+}$ glueball search in B decays. 

We are grateful to P. Kim for interesting discussions 
during his visit to UBC which 
initiated this study.

\end{document}